\catcode`\@=11
\expandafter\ifx\csname @iasmacros\endcsname\relax
	\global\let\@iasmacros=\par
\else	\endinput
\fi
\catcode`\@=12


\def\rmb{\seventeenrm}

\def\itb{icmsy8}


\def\doublespace{\baselineskip=2\normalbaselineskip}


\def\nonarrower{\advance\leftskip by-\parindent
	\advance\rightskip by-\parindent}


\def\boxit#1{\vbox{\hrule\hbox{\vrule\kern3pt
	\vbox{\kern3pt#1\kern3pt}\kern3pt\vrule}\hrule}}

\def\hence{\leavevmode\hbox{\bf .\raise5.5pt\hbox{.}.} }

\def\dalemb#1#2{{\vbox{\hrule height.#2pt
	\hbox{\vrule width.#2pt height#1pt \kern#1pt \vrule width.#2pt}
	\hrule height.#2pt}}}
\def\gtorder{\mathrel{\raise.3ex\hbox{$>$}\mkern-14mu
             \lower0.6ex\hbox{$\sim$}}}
\def\ltorder{\mathrel{\raise.3ex\hbox{$<$}\mkern-14mu
             \lower0.6ex\hbox{$\sim$}}}

\newdimen\fullhsize
\newbox\leftcolumn
\def\twoup{\hoffset=-.5in \voffset=-.25in
  \hsize=4.75in \fullhsize=10in \vsize=6.9in
  \def\fullline{\hbox to\fullhsize}
  \let\lr=L
  \output={\if L\lr
        \global\setbox\leftcolumn=\columnbox\global\let\lr=R \advancepageno
      \else \doubleformat \global\let\lr=L\fi
    \ifnum\outputpenalty>-20000 \else\dosupereject\fi}
  \def\doubleformat{\shipout\vbox{
    \fullline{\box\leftcolumn\hfil\columnbox}\advancepageno}}
  \def\columnbox{\leftline{\vbox{\makeheadline\pagebody\makefootline}}}
  \tolerance=1000 }
\catcode`\@=11					



\font\fiverm=cmr5				
\font\fivemi=cmmi5				
\font\fivesy=cmsy5				
\font\fivebf=cmbx5				

\skewchar\fivemi='177
\skewchar\fivesy='60


\font\sixrm=cmr6				
\font\sixi=cmmi6				
\font\sixsy=cmsy6				
\font\sixbf=cmbx6				

\skewchar\sixi='177
\skewchar\sixsy='60


\font\sevenrm=cmr7				
\font\seveni=cmmi7				
\font\sevensy=cmsy7				
\font\sevenit=cmti7				
\font\sevenbf=cmbx7				

\skewchar\seveni='177
\skewchar\sevensy='60


\font\eightrm=cmr8				
\font\eighti=cmmi8				
\font\eightsy=cmsy8				
\font\eightit=cmti8				
\font\eightbf=cmbx8				

\skewchar\eighti='177
\skewchar\eightsy='60


\font\ninei=cmmi9
\font\ninesy=cmsy9

\skewchar\ninei='177
\skewchar\ninesy='60


\font\tenrm=cmr10				
\font\teni=cmmi10				
\font\tensy=cmsy10				
\font\tenex=cmex10				
\font\tenit=cmti10				
\font\tensl=cmsl10				
\font\tenbf=cmbx10				
\font\tentt=cmtt10				
\font\tenss=cmss10				
\font\tensc=cmcsc10				
\font\tenbi=cmmib10				

\skewchar\teni='177
\skewchar\tenbi='177
\skewchar\tensy='60

\def\tenpoint{\ifmmode\err@badsizechange\else
	\textfont0=\tenrm \scriptfont0=\sevenrm \scriptscriptfont0=\fiverm
	\textfont1=\teni  \scriptfont1=\seveni  \scriptscriptfont1=\fivemi
	\textfont2=\tensy \scriptfont2=\sevensy \scriptscriptfont2=\fivesy
	\textfont3=\tenex \scriptfont3=\tenex   \scriptscriptfont3=\tenex
	\textfont4=\tenit \scriptfont4=\sevenit \scriptscriptfont4=\sevenit
	\textfont5=\tensl
	\textfont6=\tenbf \scriptfont6=\sevenbf \scriptscriptfont6=\fivebf
	\textfont7=\tentt
	\textfont8=\tenbi \scriptfont8=\seveni  \scriptscriptfont8=\fivemi
	\def\rm{\tenrm\fam=0 }%
	\def\it{\tenit\fam=4 }%
	\def\sl{\tensl\fam=5 }%
	\def\bf{\tenbf\fam=6 }%
	\def\tt{\tentt\fam=7 }%
	\def\ss{\tenss}%
	\def\sc{\tensc}%
	\def\bmit{\fam=8 }%
	\rm\setparameters\setbaselines\fi}


\font\twelverm=cmr12				
\font\twelvei=cmmi12				
\font\twelvesy=cmsy10	scaled\magstep1		
\font\twelveex=cmex10	scaled\magstep1		
\font\twelveit=cmti12				
\font\twelvesl=cmsl12				
\font\twelvebf=cmbx12				
\font\twelvett=cmtt12				
\font\twelvess=cmss12				
\font\twelvesc=cmcsc10	scaled\magstep1		
\font\twelvebi=cmmib10	scaled\magstep1		

\skewchar\twelvei='177
\skewchar\twelvebi='177
\skewchar\twelvesy='60

\def\twelvepoint{\ifmmode\err@badsizechange\else
	\textfont0=\twelverm \scriptfont0=\eightrm \scriptscriptfont0=\sixrm
	\textfont1=\twelvei  \scriptfont1=\eighti  \scriptscriptfont1=\sixi
	\textfont2=\twelvesy \scriptfont2=\eightsy \scriptscriptfont2=\sixsy
	\textfont3=\twelveex \scriptfont3=\tenex   \scriptscriptfont3=\tenex
	\textfont4=\twelveit \scriptfont4=\eightit \scriptscriptfont4=\sevenit
	\textfont5=\twelvesl
	\textfont6=\twelvebf \scriptfont6=\eightbf \scriptscriptfont6=\sixbf
	\textfont7=\twelvett
	\textfont8=\twelvebi \scriptfont8=\eighti  \scriptscriptfont8=\sixi
	\def\rm{\twelverm\fam=0 }%
	\def\it{\twelveit\fam=4 }%
	\def\sl{\twelvesl\fam=5 }%
	\def\bf{\twelvebf\fam=6 }%
	\def\tt{\twelvett\fam=7 }%
	\def\ss{\twelvess}%
	\def\sc{\twelvesc}%
	\def\bmit{\fam=8 }%
	\rm\setparameters\setbaselines\fi}


\font\fourteenrm=cmr10	scaled\magstep2		
\font\fourteeni=cmmi10	scaled\magstep2		
\font\fourteensy=cmsy10	scaled\magstep2		
\font\fourteenex=cmex10	scaled\magstep2		
\font\fourteenit=cmti10	scaled\magstep2		
\font\fourteensl=cmsl10	scaled\magstep2		
\font\fourteenbf=cmbx10	scaled\magstep2		
\font\fourteentt=cmtt10	scaled\magstep2		
\font\fourteenss=cmss10	scaled\magstep2		
\font\fourteensc=cmcsc10 scaled\magstep2	
\font\fourteenbi=cmmib10 scaled\magstep2	

\skewchar\fourteeni='177
\skewchar\fourteenbi='177
\skewchar\fourteensy='60

\def\fourteenpoint{\ifmmode\err@badsizechange\else
	\textfont0=\fourteenrm \scriptfont0=\tenrm \scriptscriptfont0=\sevenrm
	\textfont1=\fourteeni  \scriptfont1=\teni  \scriptscriptfont1=\seveni
	\textfont2=\fourteensy \scriptfont2=\tensy \scriptscriptfont2=\sevensy
	\textfont3=\fourteenex \scriptfont3=\tenex \scriptscriptfont3=\tenex
	\textfont4=\fourteenit \scriptfont4=\tenit \scriptscriptfont4=\sevenit
	\textfont5=\fourteensl
	\textfont6=\fourteenbf \scriptfont6=\tenbf \scriptscriptfont6=\sevenbf
	\textfont7=\fourteentt
	\textfont8=\fourteenbi \scriptfont8=\tenbi \scriptscriptfont8=\seveni
	\def\rm{\fourteenrm\fam=0 }%
	\def\it{\fourteenit\fam=4 }%
	\def\sl{\fourteensl\fam=5 }%
	\def\bf{\fourteenbf\fam=6 }%
	\def\tt{\fourteentt\fam=7}%
	\def\ss{\fourteenss}%
	\def\sc{\fourteensc}%
	\def\bmit{\fam=8 }%
	\rm\setparameters\setbaselines\fi}


\font\seventeenrm=cmr10 scaled\magstep3		


\newdimen\rp@
\newcount\@basestretchnum
\newskip\@baseskip
\newskip\headskip
\newskip\footskip


\def\setparameters{\rp@=.1em
	\headskip=24\rp@
	\footskip=\headskip
	\delimitershortfall=5\rp@
	\nulldelimiterspace=1.2\rp@
	\scriptspace=0.5\rp@
	\abovedisplayskip=10\rp@ plus3\rp@ minus5\rp@
	\belowdisplayskip=10\rp@ plus3\rp@ minus5\rp@
	\abovedisplayshortskip=5\rp@ plus2\rp@ minus4\rp@
	\belowdisplayshortskip=10\rp@ plus3\rp@ minus5\rp@
	\normallineskip=\rp@
	\lineskip=\normallineskip
	\normallineskiplimit=0pt
	\lineskiplimit=\normallineskiplimit
	\jot=3\rp@
	\setbox0=\hbox{\the\textfont3 B}\p@renwd=\wd0
	\skip\footins=12\rp@ plus3\rp@ minus3\rp@
	\skip\topins=0pt plus0pt minus0pt}


\def\setbaselines{\maxdepth=4\rp@\baselinestretch=\@basestretchnum}


\def\baselinestretch{\afterassignment\@basestretch\@basestretchnum}
\def\@basestretch{%
	\@baseskip=12\rp@ \divide\@baseskip by1000
	\normalbaselineskip=\@basestretchnum\@baseskip
	\baselineskip=\normalbaselineskip
	\bigskipamount=\the\baselineskip
		plus.25\baselineskip minus.25\baselineskip
	\medskipamount=.5\baselineskip
		plus.125\baselineskip minus.125\baselineskip
	\smallskipamount=.25\baselineskip
		plus.0625\baselineskip minus.0625\baselineskip
	\setbox\strutbox=\hbox{\vrule height.708\baselineskip
		depth.292\baselineskip width0pt }}



\def\makeheadline{\vbox to0pt{\baselinestretch=1000
	\vskip-\headskip \vskip1.5pt
	\line{\vbox to\ht\strutbox{}\the\headline}\vss}\nointerlineskip}

\def\makefootline{\baselineskip=\footskip\line{\the\footline}}

\def\big#1{{\hbox{$\left#1\vbox to8.5\rp@ {}\right.\n@space$}}}
\def\Big#1{{\hbox{$\left#1\vbox to11.5\rp@ {}\right.\n@space$}}}
\def\bigg#1{{\hbox{$\left#1\vbox to14.5\rp@ {}\right.\n@space$}}}
\def\Bigg#1{{\hbox{$\left#1\vbox to17.5\rp@ {}\right.\n@space$}}}


\mathchardef\alpha="710B
\mathchardef\beta="710C
\mathchardef\gamma="710D
\mathchardef\delta="710E
\mathchardef\epsilon="710F
\mathchardef\zeta="7110
\mathchardef\eta="7111
\mathchardef\theta="7112
\mathchardef\iota="7113
\mathchardef\kappa="7114
\mathchardef\lambda="7115
\mathchardef\mu="7116
\mathchardef\nu="7117
\mathchardef\xi="7118
\mathchardef\pi="7119
\mathchardef\rho="711A
\mathchardef\sigma="711B
\mathchardef\tau="711C
\mathchardef\upsilon="711D
\mathchardef\phi="711E
\mathchardef\chi="711F
\mathchardef\psi="7120
\mathchardef\omega="7121
\mathchardef\varepsilon="7122
\mathchardef\vartheta="7123
\mathchardef\varpi="7124
\mathchardef\varrho="7125
\mathchardef\varsigma="7126
\mathchardef\varphi="7127
\mathchardef\imath="717B
\mathchardef\jmath="717C
\mathchardef\ell="7160
\mathchardef\wp="717D
\mathchardef\partial="7140
\mathchardef\flat="715B
\mathchardef\natural="715C
\mathchardef\sharp="715D


\def\err@badsizechange{%
	\immediate\write16{--> Size change not allowed in math mode, ignored}}

\baselinestretch=1000
\tenpoint

\catcode`\@=12					

\twelvepoint
\doublespace
\font\itb=cmr10 scaled 1400
\rightline{IASSNS-HEP-96/72}
\bigskip\medskip
\centerline{\rmb Quaternionic Quantum Mechanics and}
\centerline{\rmb Noncommutative Dynamics}
\medskip
\centerline{\itb Stephen L. Adler}
\centerline{\bf Institute for Advanced Study}
\centerline{\bf Princeton, NJ 08540}
\medskip
\bigskip\bigskip

In this talk I shall first make some brief remarks on quaternionic quantum
mechanics, and then describe recent work with A.C. Millard in which we 
show that standard complex quantum field theory can arise as the statistical 
mechanics of an underlying noncommutative dynamics.

In quaternionic quantum mechanics, the Dirac transition amplitudes 
$\langle \psi | \phi \rangle$ are quaternion valued, that is, they have the 
form $r_0+r_1 i +r_2 j + r_3 k$, where $r_{0,1,2,3}$ are real numbers and 
$i,j,k$ are quaternion imaginary units obeying $i^2=j^2=k^2=-1~,~~~
ij=-ji=k,~jk=-kj=i,~ki=-ik=j$.   The Schr\"odinger equation takes the form
$$\eqalign{
{\partial | \psi \rangle \over \partial t}=&-\tilde H |\psi \rangle~,\cr
\tilde H=&-\tilde H^{\dagger}  ~.\cr
}\eqno(1)$$
A systematic study of quaternionic quantum mechanics is given in my 
recent book [1] on ``Quaternionic Quantum Mechanics and Quantum 
Fields.''   For information on how to obtain  
the book and updates on more recent work, see my web home page (http://
www.sns.ias.edu/$\sim$adler/Html/quaternionic.html).

In the book, I discuss many aspects of the generalization of standard 
complex quantum mechanics to quaternionic Hilbert space.  Let me here 
focus on the one obvious question, can quaternionic quantum mechanics 
be relevant to physics?  A key result in the book relating to this is that 
in quaternionic quantum mechanics, the $S$-matrix is always a complex 
matrix with no dependence on the quaternionic units $j,k$, for appropriate ray 
representative choices for the states in Hilbert space.  Hence the asymptotic 
dynamics for quaternionic quantum mechanics is always an effective complex 
theory, and this means that a quaternionic Hilbert space dynamics 
may play a role as an underlying dynamics for the standard model.  

The reason for my long-standing interest in quaternionic quantum mechanics
is that it offers the possibility of elegant substructure models for 
quarks and leptons along lines outlined long ago by Harari and Shupe.  
Recently [2], I  gave a simple set of rules for constructing composite 
quarks and leptons as triply occupied quaternionic quasiparticles. The 
mixed symmetry states obtained this way correspond precisely to the three 
spin 1/2 quark-lepton families of the standard model, plus one additional 
family of (possibly massive) spin 3/2 quarks.  The fact that the 
spin 1/2 state multiplicities come out right could of course be fortuitous;  
the next step in the program is to try to give a systematic derivation of 
the rules from an underlying quaternionic Hilbert space or other 
noncommutative dynamics.  

Let me turn now to the major topic of my talk, which is to sketch how 
standard complex quantum field theory can arise as the statistical mechanics 
of an underlying noncommutative dynamics.  This investigation grew out of 
my analysis [1,~3] of the problem of taking the step from quaternionic 
quantum mechanics to quaternionic quantum field theory, which requires 
a generalization beyond the canonical quantum mechanical formalism.  
To do this, I suggested a formalism that I termed ``generalized quantum 
dynamics,'' which is a noncommutative generalization of classical Lagrangian 
and Hamiltonian dynamics.  Let me describe briefly how it works, focusing 
for simplicity on the bosonic case (all results generalize properly 
when fermionic grading is included).

Let $\{ q_r \}$ be a set of noncommuting coordinates, which act as linear 
operators on an underlying real, complex, or quaternionic Hilbert space, and 
let $\{ \dot{q}_r \}$ be their time derivatives.  From these quantities 
we form a polynomial operator Lagrangian $L[\{q_r\},\{\dot{q}_r\}]$, in 
which the order of factors is significant.  Because the coordinates $\{q_r\}$ 
do not commute with one another, we cannot define a coordinate derivative 
$\delta L/ \delta q_r$.  To deal with this problem, we use the cyclic 
invariance property of the trace:  let us define the total trace Lagrangian
$${\bf L} \equiv {\bf Tr} L \equiv {\rm Re} Tr L~,\eqno(2a)$$
where Tr denotes the trace over the underlying Hilbert space and where Re 
denotes the real part.  If we vary all of the $q_r$ and $\dot{q}_r$, in the 
corresponding variation of the total trace Lagrangian we can use cyclic 
invariance of the trace to reorder factors so that all of the $\delta q_r$ 
and $\delta \dot q_r$ stand on the right, allowing us to define derivatives 
of {\bf L} with respect to the $q_r$ and $\dot q_r$ by 
$$\delta {\bf L}=\sum_r {\bf Tr} \left( {\delta {\bf L}\over \delta q_r}
\delta q_r + {\delta {\bf L}\over \delta \dot{q}_r } 
\delta \dot {q}_r \right) ~.\eqno(2b)$$

Using this definition, it is now easy to demonstrate the following properties:

\parindent=0pt
(1)  Introducing the total trace action defined by ${\bf S}=\int_{-\infty}^
{\infty} dt {\bf L}$ and requiring stationarity as expressed by 
$0=\delta {\bf S}$, implies the {\it operator} Euler-Lagrange equations
$${\delta {\bf L} \over \delta q_r} - {d\over dt} {\delta {\bf L} \over 
\delta \dot{q}_r } =0~.\eqno(3)$$

\parindent=0pt
(2)  If we now define the operator momenta $\{ p_r \}$ and total trace 
Hamiltonian {\bf H} by 
$$\eqalign{
p_r=&{\delta {\bf L} \over \delta \dot{q}_r }~,\cr
{\bf H}=&{\bf Tr}\sum_r p_r \dot{q}_r - {\bf L} ~,\cr
}\eqno(4a)$$
then the operator Euler-Lagrange equations can be rewritten as operator 
Hamilton equations 
$${\delta {\bf H} \over \delta q_r} = -\dot{p}_r~,~~~
  {\delta {\bf H} \over \delta p_r} = \dot{q}_r~.\eqno(4b)$$

\parindent=0pt
(3)  For any total trace functional ${\bf A}[\{q_r\},\{p_r\},t]$, the 
time evolution is given by 
$${d{\bf A} \over dt}={\partial{\bf A}\over \partial t}+\{{\bf A},{\bf H}\}
~,\eqno(5a)$$
where the final term is a generalized Poisson bracket of its arguments, 
defined for any 
two total trace functionals {\bf A} and {\bf B} by 
$$\{ {\bf A}, {\bf B} \}={\bf Tr} \sum_r \left( {\delta {\bf A} \over \delta 
q_r} 
{\delta {\bf B} \over \delta p_r} - {\delta {\bf B} \over \delta q_r}
{\delta {\bf A} \over \delta p_r} \right)~.\eqno(5b)$$
Applying this to the total trace Hamiltonian {\bf H}, and assuming that 
the dynamics is time translation invariant, so that $\partial {\bf H}/
\partial t=0$, we have 
$${d {\bf H} \over dt}=\{ {\bf H}, {\bf H} \}=0~,\eqno(6)$$
telling us that the total trace Hamiltonian {\bf H} is a 
constant of the motion.  Note that there is no corresponding conserved 
{\it operator} Hamiltonian.

\parindent=0pt
(4)  The generalized Poisson bracket obeys the Jacobi identity [4]
$$\{ {\bf A}, \{ {\bf B}, {\bf C} \}\}
+ \{ {\bf C}, \{ {\bf A}, {\bf B} \}\}
+ \{ {\bf B}, \{ {\bf C}, {\bf A} \}\}=0~. \eqno(7)$$
The derivation of the Jacobi identity uses only associativity of the
multiplication in the underlying Hilbert space, which can be real, complex,
or quaternionic (but not octonionic).  As a consequence of the Jacobi 
identity [5], the symplectic structure of generalized quantum dynamics is 
closely analogous to that of classical mechanics.

\parindent=0pt
(5)  Let us define the anti-self-adjoint operator $\tilde C$ by 
$$\tilde C=\sum_r [q_r, p_r] ~.\eqno(8)$$
Then when {\bf H} involves no noncommutative constants, a simple argument 
base on cyclic invariance of the trace [6] shows that $d \tilde C/ dt=0$,
so that $\tilde C$ is a conserved operator.  

\parindent=0pt
(6)  In analogy with classical mechanics, let us define a canonical 
transformation in generalized quantum dynamics by
$$\delta p_r=-{\delta {\bf G}\over q_r}~,
~~\delta q_r={\delta {\bf G}\over p_r}~,\eqno(9)$$
where {\bf G} is any total trace functional; in particular, when 
${\bf G}={\bf H}dt$, the canonical transformation corresponds to an 
infinitesimal time step.  Let us now define a phase space measure in 
operator phase space by 
$$d\mu=\prod_{r,m,n,A} d\langle m |q_r | n\rangle^A 
d \langle m| p_r | n \rangle^A~,\eqno(10)$$
where $A$ indexes the real components of the indicated matrix elements   
(i.e., $A=0$ in real Hilbert space, $A=0,1$ in complex Hilbert space, and 
$A=0,1,2,3$ in quaternionic Hilbert space).  Then one can show [6] that $d 
\mu$   
is invariant under the general canonical transformation of Eq.~(9), and 
in particular is conserved under time evolution.  

\parindent=25pt
The existence of a conserved phase space measure implies that generalized 
quantum dynamics obeys a generalized Liouville's theorem, and this in 
turn means that the methods of statistical mechanics are applicable.  The 
canonical ensemble [6] has the form 
$$\eqalign{
\rho=&Z^{-1} e^{-\tau {\bf H} -{\bf Tr} \tilde \lambda \tilde C}~,\cr
\int d\mu \rho =1 \Rightarrow& Z=\int d\mu 
e^{-\tau {\bf H} -{\bf Tr} \tilde \lambda \tilde C} ~.\cr
}\eqno(11)$$
Here $\tau$ and $\tilde \lambda$ are ensemble parameters which are 
determined by requiring that the ensemble averages $\langle {\bf H} 
\rangle_{AV}$ 
and $\langle \tilde C \rangle_{AV}$ have specified values.  For the 
corresponding form of the microcanonical ensemble, and its relation to the 
canonical ensemble, see [7].  Introducing the standard canonical form 
for an anti-self-adjoint operator, we can write 
$$ \langle \tilde C \rangle_{AV} = i_{eff} D~, \eqno(12a)$$
with $D$ real diagonal and with
$$i_{eff}^2=-1~,~~i_{eff}^{\dagger}=-i_{eff}~,~~[i_{eff},D]=0~.\eqno(12b)$$
Henceforth we shall assume that the ensemble does not favor any 
state in the underlying Hilbert space, which implies that $D$ is 
proportional to the unit operator with a proportionality 
constant that we denote by $\hbar$, so that the canonical form of Eq.~(12a) 
reads 
$$\langle \tilde C \rangle_{AV}= i_{eff} \hbar~.\eqno(12c)$$
 
\parindent=0pt
(7)  Let us now look at the Ward identities for averages of the dynamical 
variables over $\rho$, that are analogous to the 
generalized equipartition theorem in classical 
statistical mechanics (which we recall states that for $x_r$ a coordinate 
or momentum, we have $\langle x_s \partial H/\partial x_r\rangle_{AV} 
=\beta^{-1} 
\delta_{rs}$, with $\beta$ the inverse temperature in units of Boltzmann's 
constant).  A detailed calculation [6] shows that equipartition of 
the conserved operator $\tilde C$ in generalized quantum dynamics gives 
the canonical commutation relations of complex quantum field theory, with 
$i_{eff}$ playing the role of the imaginary unit and with the averages over 
$\rho$ playing the role of the Wightman functions.  In other words, 
canonical, complex quantum field theory can arise as an {\it emergent 
property} of a deterministic underlying noncommutative operator dynamics!
\bigskip\bigskip
\parindent=25pt
To conclude, both our discussion of the $S$-matrix in quaternionic quantum 
mechanics, and of the statistical properties of generalized quantum dynamics, 
show that {\it noncommutative dynamics hides itself beneath effective 
complex quantum mechanical structures}.  This means that noncommutative 
generalizations of standard quantum mechanics and quantum field theory could  
be the vehicle for an elegant solution to the problem of unifying the 
forces--one is not restricted to searching for a solution solely 
within the framework of quantum mechanics in complex Hilbert space.

This work was supported in part by the Department of Energy under
Grant \#DE--FG02--90ER40542.  I wish to acknowledge the hospitality of
the Aspen Center for Physics where part of this work was done.
\bigskip\smallskip
\centerline{\bf References}
\item{[1]}  
S. L. Adler, ``Quaternionic Quantum Mechanics and Quantum Fields'', 
Oxford, New York, 1995.
\item{[2]}
S. L. Adler, Phys. Letters {\bf B332}, 358 (1994).

\item{[3]}  
S. L. Adler, Nucl. Phys. {\bf B415}, 195 (1994).

\item{[4]}  
S. L. Adler, G. V. Bhanot, and J. D. Weckel, J. Math. Phys. {\bf 35}, 
531 (1994).

\item{[5]}  
S. L. Adler and Y.-S. Wu, Phys. Rev. {\bf D 49}, 6705 (1994).

\item{[6]}  
S. L. Adler and A. C. Millard, ``Generalized Quantum Dynamics as 
Pre-Quantum Mechanics", Nucl. Phys. {\bf B} (in press).

\item{[7]}  
S. L. Adler and L. P. Horwitz, ``Microcanonical Ensemble and Algebra 
of Conserved Generators for Generalized Quantum Dynamics'', hep-th/9606023, 
submitted to J. Math. Phys.

\bye